\documentclass[showpacs,aps,graphicx,twocolumn]{revtex4}

\usepackage{graphicx}

\begin{document}

\title{Entanglement concentration of microwave photons based on Kerr effect in circuit QED}

\author{Hao Zhang and  Haibo Wang\footnote{Corresponding author:hbwang@bnu.edu.cn} }

\address{Department of Physics, Applied Optics Beijing Area Major Laboratory, Beijing Normal University, Beijing 100875, China}

\date{\today }

\begin{abstract}
In recent years, superconducting qubits show great potential in quantum computation. Hence, microwave photons become very interesting qubits for quantum information processing assisted by superconducting quantum computation. Here, we present the first protocol for the entanglement concentration on microwave photons, resorting to the cross-Kerr effect in circuit quantum electrodynamics (QED). Two superconducting transmission line resonators (TLRs) coupled to superconducting molecule with the $N$-type level structure induce the effective cross-Kerr effect for realizing the quantum nondemolition (QND) measurement on microwave photons. With this device,  we present a two-qubit polarization parity QND detector on the photon states of the superconducting TLRs, which can be used to  concentrate efficiently the nonlocal non-maximally entangled states of microwave photons assisted by several linear microwave elements. This protocol has a high efficiency and it may be useful for solid-state quantum information processing assisted by microwave photons.
\end{abstract}

 \maketitle

\section{Introduction}\label{sec1}

Quantum entanglement plays an extremely important role in quantum communication, such as quantum teleportation \cite{CHBennet1993}, quantum dense coding \cite{CHBennet1992,superdense}, quantum key distribution \cite{AKEkert,bbm92,lixhqkdpra}, quantum secret sharing \cite{MHillery}, and quantum secure direct communication \cite{longliupra,FGDeng2003,wangchQSDC}. The maximally entangled photons are usually used to act as the information carriers in quantum communication. However, they  unavoidably interact with the environment when they are transmitted over a noisy channel or stored in artificial atomic systems, which will lead their decoherence and make them in a partially entangled state or a mixed entangled one. To accomplish the quantum communication effectively, some interesting methods are used to depress the effect of environment noise, such as faithful qubit transmission \cite{faithful}, error-rejecting with decoherence free subspaces \cite{DFS1,DFS2,DFS3,lixhqkdpra}, entanglement purification \cite{EPP1,EPP2,EPP3,EPP4,EPP5}, and entanglement concetration \cite{CHBennett1996}.

Entanglement concentration is aimed at transforming a nonlocal partially entangled pure state into a maximally entangled state and it becomes an indispensable part in long-distance quantum communication. By far, there are some interesting entanglement concentration protocols for photon systems and atomic systems. For instance, in 1996, Bennett \emph{et al.} \cite{CHBennett1996} proposed the original entanglement concentration protocol (ECP) by means of the Schmidt projection. In 2001, two ECPs were proposed for ideal entanglement sources with polarizing beam splitters \cite{ZZhao014301,TYamamoto012304}. In 2008, Sheng \emph{et al.} \cite{YBSheng2008} proposed a  repeatable ECP  and it improves the efficiency largely by iteration of the entanglement concentration process three times. In fact, the existing ECPs can be divided into two categories depending on whether the parameters of the less-entangled states are known \cite{SBose194,YBSheng012307,FGDeng022311,BCRen012302} or not \cite{CHBennett1996,ZZhao014301,TYamamoto012304,YBSheng2008}. When the parameters are known, a nonlocal photon system is enough for entanglement concentration \cite{SBose194,YBSheng012307,FGDeng022311,BCRen012302}, and it is far more efficient than those with unknown coefficients \cite{CHBennett1996,ZZhao014301,TYamamoto012304,YBSheng2008}. In 1999, Bose \emph{et al.} \cite{SBose194} proposed an ECP for partially entangled pure state with known coefficients based on the entanglement swapping of two EPR pairs.  In 2012, Sheng \emph{et al.} \cite{YBSheng012307} proposed two efficient ECPs for the less-entangled states with known parameters with the help of an ancillary single photon.  In 2013, Ren \emph{et al.} \cite{BCRen012302} presented the parameter-splitting method for the concetraton of  nonlocally partially entangled states with  known coefficients, and this fascinating method  can accomplish the concentration with the maximal success probability by executing the protocol only once, just resorting to linear-optical elements.  They \cite{BCRen012302} also presented the pioneering hyper-ECP for unknown polarization-spatial less-hyperentangled states with linear-optical elements only. In 2014, Ren and Long \cite{BCRen6547} gave a general hyper-ECP and another high-efficiency hyper-ECP \cite{BCRen16444}. Li and Ghose \cite{XHLi125201} proposed an interesting hyper-ECP with linear optics and another two efficient hyper-ECPs \cite{XHLi3550,Lixhpraecp} with  nonlinearity in 2015. Wang \emph{et al.} \cite{WangECPOE15} and Cao \emph{et al.} \cite{WangECPAP16} gave two good hyper-ECPs with or without  local entanglement resource, respectively. Also, Wang \emph{et al.} \cite{CWangECPPRA11,CWangECPPRA12} presented two interesting ECPs for electron-spin systems.

Circuit quantum electrodynamics (QED) is composed of a superconducting qubit coupled to a superconducting resonator \cite{ABlais,AWallraff}. It plays an important role in quantum information processing as it has the good scalability \cite{ABlais2,DiCarlo,LongcircuitPRA,Wangsuperconducting,circuitTianlPRL,3q,3q1,Frederick1,HuaMSR1,HuaMSR2}. Many achievements have been accomplished in circuit QED \cite{JQYou2005,LFrunzio,AAHouck,MHofheinz,JMajer,DISchuster,BRJohnson}. In recent years, Kerr effect has been generally concerned and researched in circuit QED \cite{SRebic2009,SKumarPRB2010,YHu,GKirchmair,ICHoi,ETHolland}. For example, Rebi\'{c} \emph{et al.} \cite{SRebic2009} presented the giant Kerr effect in 2009. In 2011, Hu \emph{et al.} \cite{YHu} proposed a scheme for implementing cross-Kerr effect in circuit QED. In 2013, Kirchmair \emph{et al.} \cite{GKirchmair} observed quantum state collapse and revival due to the single-photon Kerr effect and Hoi \emph{et al.} \cite{ICHoi} demonstrated the giant cross-Kerr effect for propagating microwaves experimentally. In 2015, Holland \emph{et al.} \cite{ETHolland} demonstrated single-photon resolved cross-Kerr effect between two superconducting cavities. Due to the low loss and strong anti-interference during transmission, microwave photons hold good prospects for both classical and quantum communication. In recent years, superconducting quantum computation shows great potential in quantum information processing. Combining the microwave quantum communication with quantum computation is an interesting research area. Because of dissipation in the process of microwave photon transmission and storage, the maximally entangled state can not keep all the time. Entanglement concentration of microwave-photon states is an extremely significant and necessary task. However, the research of this area has been in blank. The cross-Kerr effect can be used for quantum nondemolition (QND) measurement. Polarization is an important degree of freedom of microwave and one can manipulate it by adjusting parameters of particular materials \cite{DMPozar}, for example, anisotropic metamaterials \cite{JMHao} and photonic crystal \cite{DRSolli1}. Some polarization linear elements have been implemented for microwave photon theoretically and experimentally \cite{DRSolli1,RPTorres,MKLiu}, such as microwave polarization beam splitter (PBS) and polarization rotator.

In this paper, we propose the first physically feasible protocol to achieve the concentration on the polarization state of nonlocal entangled microwave photons in circuit QEDs. The crucial step of our protocol is the construction of a two-qubit polarization parity QND detector to check the polarization parity of two microwave photons. The QND measurement is accomplished by cross-Kerr effect and other polarization linear elements.  This protocol has a high efficiency and it maybe has good application in solid-state quantum computation assisited by microwave photons, especially the combination of quantum computation and quantum communication.

This article is organized as follows: In Sec.~\ref{basic}, we introduce the physical implementation of a superconducting molecule in Sec.~\ref{sec21} and the process of inducing cross-Kerr effect in circuit QED in Sec.~\ref{sec22}. In Sec.~\ref{sec23}, we present the scheme for making QND measurement on two cascade TLRs. In Sec.~\ref{sec3}, we design a two-qubit polarization parity QND detector and present the principle for the concentration of microwave photon pairs. A discussion and a summary are given in Sec.~\ref{sec4}.

\begin{figure}[!ht]
\begin{center}
\includegraphics[width=7.8cm,angle=0]{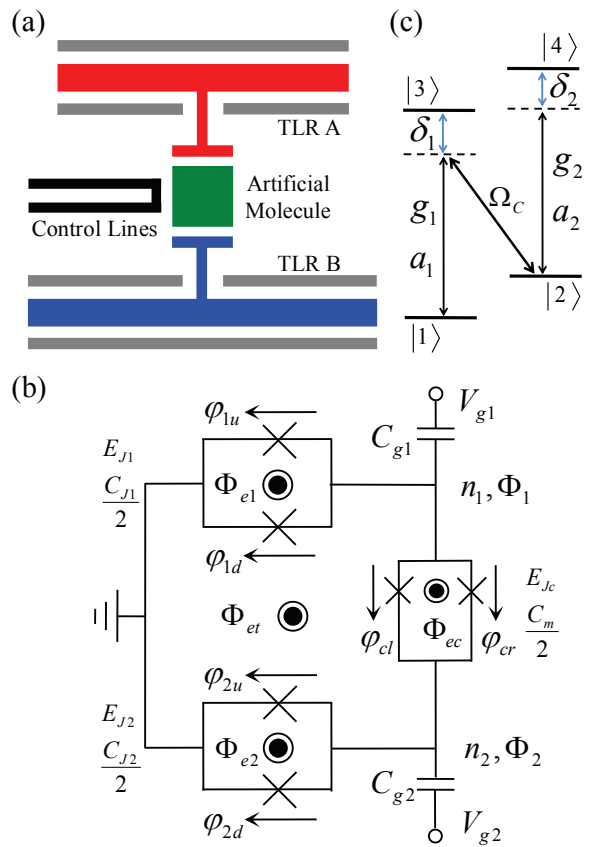}
\caption{(a) Scheme for the proposed cross-Kerr effect between two TLRs. The TLR A (upper, red) and the TLR B (lower, blue) are capacitively coupled to a superconducting molecule (middle, green). The superconducting molecule is described in Fig. \ref{fig1}(b) in detail and it can be controlled by external coils (left, black). (b) Detailed schematic circuit of the proposed superconducting molecule \cite{YHu}. The middle circuit stands for coupled transmon qubits. The crosses and the circles in the SQUID loops represent the josephson junctions and the external flux biases, respectively. (c) N-type level structure of the molecule which is described in Fig. \ref{fig1}(b).}
\label{fig1}
\end{center}
\end{figure}

\section{QND measurement of total photon number of resonators} \label{basic}

\subsection{Cross-Kerr effect in circuit QED}
\label{sec21}

The schematic diagram for realizing the cross-Kerr effect between the microwave photons in the two TLRs capacitively coupled to an artificial superconducting molecule is shown in Fig. \ref{fig1}(a). Two color bars represent two TLRs and the square on the middle is the superconducting molecule whose structure and energy levels are depicted in Fig. \ref{fig1}(b) and (c), respectively. The control line on the left is used to adjust the character of the superconducting molecule which is made up of two superconducting transmon qubits \cite{JKoch} and a superconducting quantum interference device (SQUID) \cite{JSiewert}. The two qubits couple to each other through the SQUID.
The two loops on the left and one loop on the right stand for two transmon qubits and SQUID, respectively. Crosses and circles of every loop represent the Josephson junctions and the external flux biases, respectively. The Josephson junctions in each loop are identical. $E_{Ji}$ $(i=c,1,2)$ and $C_{j}/2$ $(j=m,1,2)$ represent the energy and capacitance of Josephson junctions, respectively. External fluxes labeled with $\Phi_{e1}$, $\Phi_{e2}$, and $\Phi_{ec}$ are applied to SQUID loops of the two qubits and coupling SQUID, respectively. The additional external flux $\Phi_{et}$ is applied to the center loop of the molecule. Each loop has two arrows on their both sides and they stand for the gauge-invariant phases across the Josephson junctions, marked with $\varphi_{cr}$, $\varphi_{cl}$, $\varphi_{1u}$, $\varphi_{1d}$, $\varphi_{2u}$, and $\varphi_{2d}$, respectively. $V_{\text{g}1}$ and $V_{\text{g}2}$ near the capacitors are the gate voltages which bias the two corresponding qubits via the gate capacitors $C_{\text{g}1}$ and $C_{\text{g}2}$, respectively.

Under the condition of fluxoid quantization, the Hamiltonian of the superconducting molecule is given by \cite{YHu}
\begin{eqnarray}        \label{eq1}
H_{0}\!\!&=&\!\!\!\!\sum_{i=1,2}[4E_{ci}(n_{i}\!-\!n_{gi})^{2}\!-\!2E_{Ji}\cos(\pi\Phi_{ei}/\Phi_{0})\!\cos\Phi_{i}]\nonumber\\
\!\!\!\!\!\!\!\!&&+4E_{m}(n_{1}\!-\!n_{g1})(n_{2}\!-\!n_{g2})\!-\!E_{Jm}\cos(\Phi\!+\!\phi),\;\;\;\;\;\;\;\;
\end{eqnarray}
where $E_{c1,2}=e^{2}C_{\Sigma2,1}/[2(C_{\Sigma1}C_{\Sigma2}-C_{m}^{2})]$ are the effective Cooper-pair charging energies and $C_{\Sigma i}=C_{gi}+C_{Ji}+C_{m}$ is the sum of all  capacitances around the $i$-th qubit. $n_{i}$ $(i=1,2)$ denotes the canonical conjugate variable to the phase of the superconducting islands $\Phi_{i}$ and $\Phi_{i}=(\varphi_{iu}+\varphi_{id})/2$. $n_{gi}=C_{gi}V_{gi}/2e$ is the gate-induced charge number. $\Phi_{0}=h/2e$ is the flux quanta. $E_{m}=e^{2}C_{m}/(C_{\Sigma1}C_{\Sigma2}-C_{m}^{2})$ and $E_{Jm}=2E_{Jc}\cos(\pi\Phi_{ec}/\Phi_{0})$ are the capacitive coupling strength between the transmon qubits and the tunable effective Josephson tunnel energy of the coupling SQUID, respectively. $\Phi=\Phi_{1}-\Phi_{2}$ and $\phi=\pi(\Phi_{e1}+\Phi_{e2}+\Phi_{ec}+2\Phi_{et})/\Phi_{0}$. Here we assume that $\phi\equiv 0$ and two transmon qubits are identical, and then $2E_{J1}\cos(\pi\Phi_{e1}/\Phi_{0})=2E_{J2}\cos(\pi\Phi_{e2}/\Phi_{0})=E_{J}$, $E_{c1}=E_{c2}=E_{c}$, $C_{J1}=C_{J2}=C_{J}$, and $C_{\Sigma1}=C_{\Sigma2}=C_{\Sigma}$. One can utilize the two-level language in the region $E_{J}\ll E_{c}$ to get the $N$-type level form \cite{JKoch}. The eigenstates read \cite{YHu}
\begin{eqnarray}        \label{eq2}
|1\rangle&=&\cos\theta|\uparrow\uparrow\rangle-\sin\theta|\downarrow\downarrow\rangle,\nonumber\\
|2\rangle&=&(|\uparrow\downarrow\rangle-|\downarrow\uparrow\rangle)/\sqrt{2},\nonumber\\
|3\rangle&=&(|\uparrow\downarrow\rangle+|\downarrow\uparrow\rangle)/\sqrt{2},\nonumber\\
|4\rangle&=&\sin\theta|\uparrow\uparrow\rangle+\cos\theta|\downarrow\downarrow\rangle,
\end{eqnarray}
and the corresponding eigenvalues are given as
\begin{eqnarray}        \label{eq3}
E_{1}&=&-E_{m1}-(\omega^{2}+E_{m-}^{2})^{1/2},\nonumber\\
E_{2}&=&E_{m1}-E_{m+},\nonumber\\
E_{3}&=&E_{m1}+E_{m+},\nonumber\\
E_{4}&=&-E_{m1}+(\omega^{2}+E_{m-}^{2})^{1/2},
\end{eqnarray}
with the symbols $E_{m1}=$exp$(-\alpha^{2})E_{Jm}\alpha^{4}/4$, $E_{m\pm}=E_{Jm}\alpha^{2}$exp$(-\alpha^{2})\pm E_{m}\alpha^{-2}$, and $\theta=[\arctan(E_{m-}/\omega)]/2$. $\alpha=(2E_{c}/E_{J})^{1/4}$ and $\omega=(8E_{c}E_{J})^{1/2}-E_{c}$.  One can modify the rough energy scale of the molecule and the energy structure by tuning $E_{J}$ and $E_{Jm}$, shown in Fig. \ref{fig1}(c).

The Hamiltonian of the system composed of the TLRs A and B is (with $\hbar=1$)
\begin{eqnarray}        \label{eq4}
H_{TLRs}=\omega_{A}a^{\dag}a+\omega_{B}b^{\dag}b,
\end{eqnarray}
where $\omega_{A,B}=2\pi/(L_{A,B}\sqrt{Fc})$ and $L_{A,B}$ is the length of corresponding TLR. $F$ and $c$ are the inductance and capacitance of the TLRs per unit length, respectively. $a$ $(a^{\dag})$ and $b$ $(b^{\dag})$ are the annihilation (creation) operators of two TLR modes, respectively. $C_{Aj}$ ($C_{Bj}$) with $j=1,2$ is the coupling capacitance between  TLR A (B) and the $j$-th qubit. Their magnitudes are set to be $C_{A1}=C_{A2}$ and $C_{B1}=C_{B2}$. The locations of these capacitors are $x_{A1}=x_{A2}=0$ and  $x_{B1}=L_{B}/2-x_{B2}=L_{B}/8$. We add a classical pump field to connect the energy level between $|2\rangle$ and $|3\rangle$ and the form is expressed with \cite{YHu}
\begin{eqnarray}        \label{eq5}
H_{p}=-\Omega_{c}\left[\exp(i\omega_{p}t)|2\rangle\langle3|+\exp(-i\omega_{p}t)|3\rangle\langle2|\right],
\end{eqnarray}
where $\Omega_{c}$ and $\omega_{p}$ are the intensity and frequency of classical pump field. When the classical pump field is tuned in dark resonance with  TLR A, the Hamiltonian can be expressed with the form
\begin{eqnarray}        \label{eq6}
H_{cs}&=&i\text{g}_{A1}(\sigma_{13}a^{\dag}-\sigma_{31}a)+i\text{g}_{B2}(\sigma_{24}b^{\dag}-\sigma_{42}b)\nonumber\\
&&-\Omega_{c}[exp(i\omega_{p}t)|2\rangle\langle3|+exp(-i\omega_{p}t)|3\rangle\langle2|],\;\;\;\;\;\;
\end{eqnarray}
where $\text{g}_{A1}$ and $\text{g}_{B2}$ are the coupling factors \cite{YHu}. In the interaction picture, the Hamiltonian of the whole system composed of the two TLRs and the superconducting molecule is \cite{YHu}
\begin{eqnarray}        \label{eq7}
H_{sys}&=&\delta_{1}\sigma_{33}+\delta_{2}\sigma_{44}+i\text{g}_{A1}(\sigma_{13}a^{\dag}-\sigma_{31}a)\nonumber\\
&&+i\text{g}_{B2}(\sigma_{24}b^{\dag}-\sigma_{42}b)-\Omega_{c}(\sigma_{23}-\sigma_{32}).\;\;\;\;\;\;
\end{eqnarray}
Here $\sigma_{ij}=|i\rangle\langle j|$.
With the limit that $\mid\text{g}_{A1}/\Omega_{c}\mid^{2}\ll 1$, $\mid\text{g}_{B2}\mid\ll \mid\delta_{2}\mid$ \cite{AImamoglu1997}, and adiabatically eliminating the atomic degrees of freedom, one obtain the ultima effective cross-Kerr interaction Hamiltonian  \cite{YHu}
\begin{eqnarray}        \label{eq8}
H_{eff}=-\frac{\text{g}_{1}^{2}\text{g}_{2}^{2}}{\delta_{2}\Omega^{2}_{c}}a^{\dag}ab^{\dag}b,
\end{eqnarray}
where the detuning $\delta_{2}=E_{42}-\omega_{B}$ and cross-Kerr coefficient $\chi=-\text{g}_{1}^{2}\text{g}_{2}^{2}/(\delta_{2}\Omega^{2}_{c})$.  Here we neglect the subscript letters A and B for convenience.

\begin{figure}[!ht]
\begin{center}
\includegraphics[width=8.0cm,angle=0]{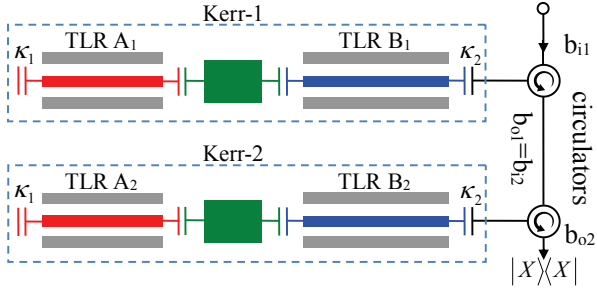}
\caption{Schematic diagram for the QND measurement on the joint photon number states of two TLRs $A_1$ and $A_2$. Both the TLRs $A_1$ and $A_2$ with the decay rate $\kappa_{1}$ are the storage resonators and both the TLR $B_1$ and $B_2$ with the decay rate $\kappa_{2}$ are the readout resonators, respectively. The element  which labels with a circular arrow in a big circle is a circulator. The direction of arrow stands for  the spread direction of the signal.}
\label{fig2}
\end{center}
\end{figure}

\subsection{Physical implementation for the  QND measurement on TLRs based on cross-Kerr effect}
\label{sec22}

Here, we present a scheme  for the QND measurement on the photon number of the cascade TLR $A_1$ and TLR $A_2$. The schematic diagram is depicted in Fig.~\ref{fig2}. It is realized based on the cross-Kerr effect in circuit QED. The Hamiltonian for each cross-Kerr medium is assumed to be
\begin{eqnarray}        \label{eq9}
H=\chi a^{\dag}a\,b^{\dag}b=\chi\, \hat{n}_{a}b^{\dag}b,
\end{eqnarray}
where the cross-Kerr coefficient $\chi=-\text{g}_{1}^{2}\text{g}_{2}^{2}/(\delta_{2}\Omega^{2}_{c})$.
$a^{\dag}$ ($a$) and $b^{\dag}$ ($b$) are the creation (annihilation) operators of TLR $A_i$ and TLR $B_i$ ($i=1,2$), respectively. The Heisenberg-Langevin equations for the two cross-Kerr medium are expressed with
\begin{eqnarray}        \label{eq10}
\dot{b_{j}}=-i\chi\, n_{j}b_{j}-\frac{\kappa_{2}}{2}b_{j}-\sqrt{\kappa_{2}}\,b^{in}_{j},
\end{eqnarray}
where $j=1$, $2$, and $n_{j}$ stands for the photon number of the $j$-th readout resonator. We assume that the decay rates of $B_{1}$ and $B_{2}$ labeled with $\kappa_{2}$ are the same ones. In this system, we consider the decay rate $\kappa_{2}\gg \chi\langle n_{j}\rangle$. The relationship between the output field and the input field is a standard cavity input-output equation $b_{out}=b_{in}+\sqrt{\kappa_{2}}\,b$ \cite{DFWalls}. $b_{in}$ and $b^{\dag}_{in}$ satisfy the standard commutation relations $[b_{in}(t),b^{\dag}_{in}(t^{\prime})]=\delta(t-t^{\prime})$. We derive the reflection coefficients which are given by
\begin{eqnarray}        \label{eq11}
r_{j}=\frac{b^{out}_{j}}{b^{in}_{j}}=\frac{i\chi n_{j}-\frac{\kappa_{2}}{2}}{i\chi n_{j}+\frac{\kappa_{2}}{2}}.
\end{eqnarray}

First, let us consider the case in which TLR $A_{1}$ is in the Fock state $|n_{1}\rangle$. We input a probe field in the coherent state $|\alpha\rangle=D(\alpha)|0\rangle_{1}^{in}=exp(\alpha a^{\dag}_{in}-\alpha^{\ast} a_{in})|0\rangle_{1}^{in}$ to interact
with the cross-Kerr system through TLR $B_{1}$. The state of the system composed of the signal photons and the probe light will evolve from $|n_{1}\rangle\vert \alpha\rangle$ to
\begin{eqnarray}        \label{eq12}
|\psi_{1}\rangle=|n_{1}\rangle D(\alpha/r_{1})|0\rangle_{1}^{out}=|n_{1}\rangle |e^{i\theta_{n1}}\alpha\rangle_{1}^{out},
\end{eqnarray}
where $\theta_{n1}= $ arg$(\frac{1}{r_{1}})$ which depends on the photon number $n_{1}$. If one measures this phase shift via an $X$ homodyne measurement, he can infer the Fock state $|n_{1}\rangle$.

Now, let us analyze joint Fock state of the whole cascaded system with cross-Kerr effect. Here the input field of resonator $B_{2}$ is just the output field of resonator $B_{1}$, that is,  $b^{in}_{2}=b^{out}_{1}$. Therefore the output of resonator $B_{2}$ connects with input of $B_{1}$ through $b^{out}_{2}=r_{2}r_{1}b^{in}_{1}$. When the coherent state $|\alpha\rangle_{p}$ passes through the TLRs $B_{1}$ and $B_{2}$, its state becomes
\begin{eqnarray}        \label{eq13}
|\alpha\rangle_{2}^{out}=D(\frac{\alpha}{r_{2}r_{1}})|0\rangle_{2}^{out}=|e^{i\theta_{n1+n2}}\alpha\rangle_{2}^{out},
\end{eqnarray}
where
\begin{eqnarray}        \label{eq14}
\theta_{n1+n2}= \text{arg}(\frac{1}{r_{2}r_{1}})= \text{arg}(\frac{1}{r_{1}})+\text{arg}(\frac{1}{r_{2}}).
\end{eqnarray}
According to the magnitude of total phase shift $\theta_{n1+n2}$ which depend on $n_{1}$ and $n_{2}$ simultaneously, one can distinguish the different joint Fock states of the system composed of the TLRs $A_{1}$ and $A_{2}$.

\begin{figure}[!ht]
\begin{center}
\includegraphics[width=7.2cm,angle=0]{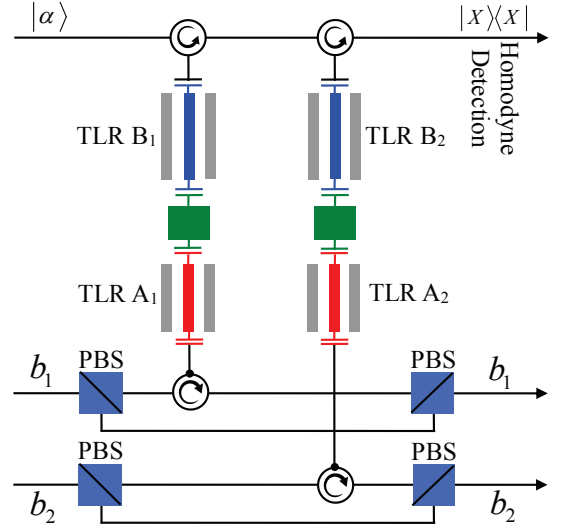}
\caption{Schematic diagram for the QND measurement on the polarization parity of microwave photons in two TLRs. The pattern which consists of a circular arrow and a circle represents a circulator. PBS is a polarizing beam splitter for microwave photons.}
\label{fig3}
\end{center}
\end{figure}

\subsection{Construction of the polarization parity QND detector on the microwave photons in two TLRs}
\label{sec23}

The principle of our polarization parity QND detector on  the microwave photons in two TLRs is shown in Fig.~\ref{fig3}. Let us assume that the cross-Kerr effects in the two systems are the same ones. In our scheme for the polarization parity detector, we add several polarization engineering elements, including the microwave PBSs \cite{DRSolli1} and polarization rotators \cite{RPTorres,MKLiu}, in microwave photon propagation path. We assume the $|H\rangle$ state is parallel to TLRs. $b_{1}$ and $b_{2}$ stand for the up spatial mode and the down spatial mode, respectively. When the microwave photons from  $b_{1}$ and $b_{2}$ are prepared in the state $|HV\rangle$, there will be a phase shift $\theta_{1}$ on the coherent state of the probe light after it passes through the left TLR, but the right TLR makes no difference on the coherent state. On the contrary, the left superconducting system makes no difference on the coherent state and the right one results in a phase shift $\theta_{1}$ if the signal photons are in the state $|VH\rangle$. Those two situations have the same result and they are undistinguishable in this QND measurement system. The states $|HH\rangle$ and $|VV\rangle$ result in the phase shifts $\theta_{2}$ and $\theta_{0}$, respectively. With an $X$ homodyne measurement on the probe light, one will get the relation between the phase shifts and the states of the signal light, shown in  TABLE \ref{Tab1}. This is a two-qubit polarization parity QND detector on the microwave photons in two TLRs. We call the same polarization states the even-parity ones and the different polarization states the odd-parity ones. One can distinguish the parity of states of two signal photons, according to the result of the detector.

\begin{table}
  \begin{center}
  \caption{Corresponding relation between the phase shifts and the states of the signal light. The range of coefficients $a$ and $b$ are $[0,1]$. But $a$ and $b$ cannot be zero simultaneously.}
    \begin{tabular}{cccccccccccc}
      \hline\hline
      Phase shift &&&&&&&&&&& State $|b_{1}b_{2}\rangle$  \\ \hline
      $\theta_{0}$ &&&&&&&&&&&$|VV\rangle$  \\ 
      $\theta_{1}$ &&&&&&&&&&& $a|HV\rangle+b|VH\rangle$ \\ 
      $\theta_{2}$ &&&&&&&&&&& $|HH\rangle$ \\ \hline\hline
    \end{tabular}\label{Tab1}
  \end{center}
\end{table}

\begin{figure}[!ht]
\begin{center}
\includegraphics[width=8.0cm,angle=0]{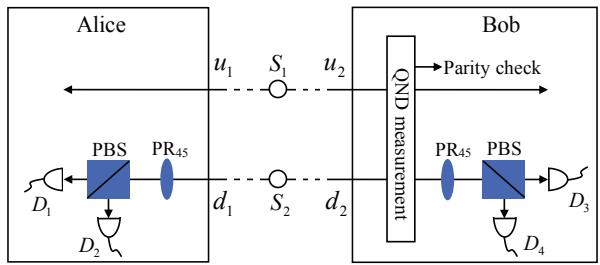}
\caption{Schematic diagram of our protocol for the nonlocal concentration on partially entangled microwave photon pairs with polarization parity-check QNDs based on cross-Kerr effect. $S_{1}$ and $S_{2}$ are the two identical ideal entanglement sources for microwave photons. $PR_{45}$  represents a polarization rotator  which can rotate the direction of polarization with $45^{\circ}$.  $D_i$ ($i=1,2,3,4$) is a microwave-photon detector.}
\label{fig4}
\end{center}
\end{figure}

\section{Nonlocal concentration on the polarization of partially entangled microwave-photon pairs in TLRs}
\label{sec3}

The principle of our protocol for the  concentration on the polarization of nonlocal partially entangled microwave-photon pairs in TLRs is shown in Fig. \ref{fig4}. Suppose that there are two microwave-photon pairs $u_1u_2$ and $d_1d_2$ shared by the two parties in quantum communication, say Alice and Bob. The photon pairs are in the partially entangled state
\begin{eqnarray}        \label{eq15}
|\psi\rangle_{u_{1}u_{2}}&=&x|H\rangle_{u_{1}}|H\rangle_{u_{2}}+y|V\rangle_{u_{1}}|V\rangle_{u_{2}},\nonumber\\
\nonumber\\
|\psi\rangle_{d_{1}d_{2}}&=&x|H\rangle_{d_{1}}|H\rangle_{d_{2}}+y|V\rangle_{d_{1}}|V\rangle_{d_{2}},
\end{eqnarray}
where $|x|^{2}+|y|^{2}=1$. The state of the system composed of the four photons is described as follow
\begin{eqnarray}        \label{eq16}
|\psi\rangle_{1}&=&|\psi\rangle_{u_{1}u_{2}}\otimes|\psi\rangle_{d_{1}d_{2}}\nonumber\\
&=&x^{2}|H\rangle_{u_{1}}|H\rangle_{u_{2}}|H\rangle_{d_{1}}|H\rangle_{d_{2}}\nonumber\\
&&+xy|H\rangle_{u_{1}}|H\rangle_{u_{2}}|V\rangle_{d_{1}}|V\rangle_{d_{2}}\nonumber\\
&&+xy|V\rangle_{u_{1}}|V\rangle_{u_{2}}|H\rangle_{d_{1}}|H\rangle_{d_{2}}\nonumber\\
&&+y^{2}|V\rangle_{u_{1}}|V\rangle_{u_{2}}|V\rangle_{d_{1}}|V\rangle_{d_{2}}.
\end{eqnarray}
Now Bob sends the microwave photons $u_{2}$ and $d_{2}$ into the two-qubit polarization parity QND detector shown in Fig. \ref{fig3}. Bob can get the state of the system composed of the microwave photons and the probe light as
\begin{eqnarray}        \label{eq17}
|\psi\rangle_{3}&=&x^{2}|H\rangle_{u_{1}}|H\rangle_{d_{1}}|H\rangle_{u_{2}}|H\rangle_{d_{2}}|\alpha e^{i\theta_{2}}\rangle_{p}\nonumber\\
&&+xy(|H\rangle_{u_{1}}|V\rangle_{d_{1}}|H\rangle_{u_{2}}|V\rangle_{d_{2}}\nonumber\\
&&+|V\rangle_{u_{1}}|H\rangle_{d_{1}}|V\rangle_{u_{2}}|H\rangle_{d_{2}})|\alpha e^{i\theta_{1}}\rangle_{p}\nonumber\\
&&+y^{2}|V\rangle_{u_{1}}|V\rangle_{d_{1}}|V\rangle_{u_{2}}|V\rangle_{d_{2}}|\alpha e^{i\theta_{0}}\rangle_{p}.
\end{eqnarray}
This equation indicates Bob will get three possible results about the phase shift of the coherent state. When the result of the homodyne detection is $\theta_{1}$, Bob tells Alice to keep those microwave photon pairs immediately. If not, those pairs are discarded. This QND measurement is a process of parity check measurement. After this operation, the state $|\psi\rangle_{3}$ is transformed to $|\psi\rangle_{4}$ with probability $P=2|xy|^{2}$. Here
\begin{eqnarray}        \label{eq18}
|\psi\rangle_{4}&=&\frac{1}{\sqrt{2}}(|H\rangle_{u_{1}}|H\rangle_{u_{2}}|V\rangle_{d_{1}}|V\rangle_{d_{2}}\nonumber\\
&&+|V\rangle_{u_{1}}|V\rangle_{u_{2}}|H\rangle_{d_{1}}|H\rangle_{d_{2}}).
\end{eqnarray}

Now, Alice and Bob make  $d_{1}$ and $d_{2}$ pass through a $45^{\circ}$ microwave polarization rotator, respectively. The transformations of this rotation on a microwave photon are given by
\begin{eqnarray}        \label{eq19}
|H\rangle_{d_{i}}&\rightarrow&\frac{1}{\sqrt{2}}(|H\rangle_{d_{i}}+|V\rangle_{d_{i}}),\nonumber\\
\nonumber\\
|V\rangle_{d_{i}}&\rightarrow&\frac{1}{\sqrt{2}}(|H\rangle_{d_{i}}-|V\rangle_{d_{i}}),
\end{eqnarray}
where $i=3,4$.
Hence, after the $45^{\circ}$ rotation, the state $|\psi\rangle_{4}$ will turn to
\begin{eqnarray}\label{eq20}
|\psi'\rangle_{4}&=&\frac{1}{2\sqrt{2}}(|H\rangle_{u_{1}}|H\rangle_{u_{2}}+|V\rangle_{u_{1}}|V\rangle_{u_{2}})\nonumber\\
&&\otimes(|H\rangle_{d_{1}}|H\rangle_{d_{2}}+|V\rangle_{d_{1}}|V\rangle_{d_{2}})\nonumber\\
&&+\frac{1}{2\sqrt{2}}(|V\rangle_{u_{1}}|V\rangle_{u_{2}}-|H\rangle_{u_{1}}|H\rangle_{u_{2}})\nonumber\\
&&\otimes(|H\rangle_{d_{1}}|V\rangle_{d_{2}}+|V\rangle_{d_{1}}|H\rangle_{d_{2}}).
\end{eqnarray}
At last, Alice and Bob use a microwave PBS to pass through $|H\rangle$ and reflect $|V\rangle$ photons, respectively. Combining Eq. (\ref{eq20}) and Fig. \ref{fig4}, one can find that if the detectors $D_{1}$ and $D_{3}$ or $D_{2}$ and $D_{4}$ are in response, Alice and Bob can get the state the two microwave photons from the up spatial modes $u_{1}u_{2}$ with the form
\begin{eqnarray}        \label{eq21}
|\Psi^{+}\rangle_{u_{1}u_{2}}=\frac{1}{\sqrt{2}}(|H\rangle_{u_{1}}|H\rangle_{u_{2}}+|V\rangle_{u_{1}}|V\rangle_{u_{2}}).
\end{eqnarray}
If the responsive detectors are $D_{1}$ and $D_{4}$ or $D_{2}$ and $D_{3}$, the state of the two microwave photons from the up spatial modes $u_{1}u_{2}$ becomes
\begin{eqnarray}        \label{eq22}
|\Psi^{-}\rangle_{u_{1}u_{2}}=\frac{1}{\sqrt{2}}(|V\rangle_{u_{1}}|V\rangle_{u_{2}}-|H\rangle_{u_{1}}|H\rangle_{u_{2}}).
\end{eqnarray}
After Alice or Bob  rotates the phase on his or her photon, they will get the maximally entangled state  $|\Psi^{+}\rangle$.

\begin{figure}[!ht]
\begin{center}
\includegraphics[width=8.0cm,angle=0]{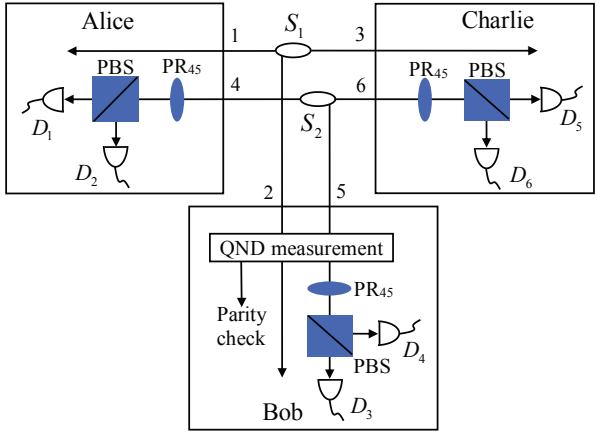}
\caption{Schematic of making a concentration on less entangled tripartite GHZ microwave photon states. The process of this protocol is similar to the situation with two users. The meanings of marks of elements are the same as those in Fig. \ref{fig4}. $S_{1}$ and $S_{2}$ represent two identical ideal tripartite entanglement microwave photon sources.}
\label{fig5}
\end{center}
\end{figure}

\section{discussion and summary}
\label{sec4}

Our ECP could be extended to the multi-user situation. For a less entangled tripartite Greenberg-Horne-Zeilinger (GHZ) type microwave photon state, we suppose it has the form
\begin{eqnarray}        \label{eq23}
|\Phi^{+}\rangle_{1}&=&x|H\rangle_{1}|H\rangle_{2}|H\rangle_{3}+y|V\rangle_{1}|V\rangle_{2}|V\rangle_{3},\nonumber\\
\nonumber\\
|\Phi^{+}\rangle_{2}&=&x|H\rangle_{4}|H\rangle_{5}|H\rangle_{6}+y|V\rangle_{4}|V\rangle_{5}|V\rangle_{6},
\end{eqnarray}
where $|x|^{2}+|y|^{2}=1$. 1 and 4 belong to Alice, 2 and 5 belong to Bob, 3 and 6 belong to Charlie.
As shown in Fig. \ref{fig5}, the state of the composite system composed of two three-photon subsystems can be written as
\begin{eqnarray}        \label{eq24}
|\Phi\rangle_{1}&=&|\Phi^{+}\rangle_{1}\otimes|\Phi^{+}\rangle_{2}\nonumber\\
&=&x^{2}|H\rangle_{1}|H\rangle_{2}|H\rangle_{3}|H\rangle_{4}|H\rangle_{5}|H\rangle_{6}\nonumber\\
&&+xy|H\rangle_{1}|H\rangle_{2}|H\rangle_{3}|V\rangle_{4}|V\rangle_{5}|V\rangle_{6}\nonumber\\
&&+xy|V\rangle_{1}|V\rangle_{2}|V\rangle_{3}|H\rangle_{4}|H\rangle_{5}|H\rangle_{6}\nonumber\\
&&+y^{2}|V\rangle_{1}|V\rangle_{2}|V\rangle_{3}|V\rangle_{4}|V\rangle_{5}|V\rangle_{6}.
\end{eqnarray}
Now Bob checks the parity of his two polarization microwave photons with his two-qubit polarization parity QND detector. The state of the composite system composed of the six microwave photons and the probe light will evolve to
\begin{eqnarray}        \label{eq25}
|\Phi\rangle_{3}&\!\!=\!\!&x^{2}|H\rangle_{1}|H\rangle_{3}|H\rangle_{4}|H\rangle_{6}|H\rangle_{2}|H\rangle_{5}|\alpha e^{i\theta_{2}}\rangle_{p}\nonumber\\
&&+xy(|H\rangle_{1}|H\rangle_{3}|V\rangle_{4}|V\rangle_{6}|H\rangle_{2}|V\rangle_{5}\nonumber\\
&&+|V\rangle_{1}|V\rangle_{3}|H\rangle_{4}|H\rangle_{6}|V\rangle_{2}|H\rangle_{5})|\alpha e^{i\theta_{1}}\rangle_{p}\nonumber\\
&&+y^{2}|V\rangle_{1}|V\rangle_{3}|V\rangle_{4}|V\rangle_{6}|V\rangle_{2}|V\rangle_{5}|\alpha e^{i\theta_{0}}\rangle_{p}.\;\;
\end{eqnarray}
Bob makes an $X$ homodyne measurement and tells Alice and Charlie to keep their photons when the result is $\theta_{1}$. Otherwise, they discard the photons. When Bob obtains the phase shift  $\theta_{1}$, the state of the six-photon system collapses to
\begin{eqnarray}        \label{eq26}
|\Phi\rangle_{4}&=&\frac{1}{\sqrt{2}}(|H\rangle_{1}|H\rangle_{2}|H\rangle_{3}|V\rangle_{4}|V\rangle_{5}|V\rangle_{6}\nonumber\\
&&+|V\rangle_{1}|V\rangle_{2}|V\rangle_{3}|H\rangle_{4}|H\rangle_{5}|H\rangle_{6})
\end{eqnarray}
with the probability $2|xy|^{2}$. After all the parties rotate their second photon by $45^{\circ}$ with their microwave polarization rotators, the state becomes
\begin{eqnarray}        \label{eq27}
|\Phi\rangle_{5}&=&(\frac{1}{\sqrt{2}})^{3}[|H\rangle_{1}|H\rangle_{2}|H\rangle_{3}(|H\rangle_{4}-|V\rangle_{4})\nonumber\\
&&(|H\rangle_{5}-|V\rangle_{5})(|H\rangle_{6}-|V\rangle_{6})\nonumber\\
&&+|V\rangle_{1}|V\rangle_{2}|V\rangle_{3}(|H\rangle_{4}+|V\rangle_{4})\nonumber\\
&&(|H\rangle_{5}+|V\rangle_{5})(|H\rangle_{6}+|V\rangle_{6})].
\end{eqnarray}
After the second photons pass through the microwave PBSs and obtain an even number of $|V\rangle$ with measurement, the state of the three-photon system 123 becomes
\begin{eqnarray}        \label{eq28}
|\Phi^{+}\rangle&=&\frac{1}{\sqrt{2}}(|H\rangle_{1}|H\rangle_{2}|H\rangle_{3}+|V\rangle_{1}|V\rangle_{2}|V\rangle_{3}).
\end{eqnarray}
If the number of $|V\rangle$  is odd, the state is given by
\begin{eqnarray}        \label{eq29}
|\Phi^{-}\rangle&=&\frac{1}{\sqrt{2}}(|V\rangle_{1}|V\rangle_{2}|V\rangle_{3}-|H\rangle_{1}|H\rangle_{2}|H\rangle_{3}).
\end{eqnarray}
Similar to above process, one can extend this concentration protocol to the situation of $N$-photon less entangled states.

Now, let us discuss the feasibility of our QND on microwave photons.
According to the protocol proposed by Hu \emph{et al}. \cite{YHu},
when the parameters of the coupled transmon qubits are set with
$E_{c}/2\pi=0.5$ GHz, $E_{J}/2\pi=16$ GHz and $E_{m}/2\pi=0.2$ GHz.
When $E_{Jm}/E_{J}$ changes from 0 to 1, the level spacing $E_{31}$,
$E_{32}$, and $E_{31}-E_{42}$ are proportional to $E_{Jm}/E_{J}$ and
their value ranges are about $8\sim 12$ GHz, $1\sim 7$ GHz, and
$0\sim 1$ GHz, respectively. The coupling strength between TLRs and
superconducting molecule could be set with $\text{g}_{1}/2\pi\sim
\text{g}_{2}/2\pi\sim 300$ MHz \cite{JMajer}. In order to satisfy
the adiabatical conditions $\mid\text{g}_{1}/\Omega_{c}\mid^{2}\ll
1$ and $\mid\text{g}_{2}\mid\ll \mid\delta_{2}\mid$, the detuning
$\delta_{2}$ and the classical pump strength $\Omega_{c}$ are
designed with $\delta_{2}/2\pi\sim\Omega_{c}/2\pi\sim 1.5$ GHz
\cite{YHu}. Hence the cross-Kerr coefficient $|\chi|/2\pi\sim 2.4$
MHz. We set $\kappa_{1}^{-1}\sim20$ $\mu$s and
$\kappa_{2}^{-1}\sim10$ ns. Large $\kappa_{2}$ can promise the probe
light pass through the resonator quickly and low $\kappa_{1}$ can
keep the signal photon in resonator for enough long time. Recent
experiments have demonstrated the giant cross-Kerr shift in circuit
QED experimentally. For example, in 2013, Hoi \emph{et al}.
\cite{ICHoi} observed the average cross-Kerr phase shifts of up to
20 degrees per photon with both coherent microwave fields at the
single-photon level by using a transmon qubit embedded in a
one-dimensional open transmission line. In 2015, Holland \emph{et
al}. \cite{ETHolland} inferred a state dependent shift
$\chi_{sc}/2\pi=2.59\pm 0.06$ MHz  according to the experimental
parameters and observed the single-photon-resolved cross-Kerr
interaction between the two three-dimensional cavities which
connected by transmon qubit.

In summary, we have proposed the first protocol to concentrate the
quantum state of nonlocal microwave photons. The protocol aims at
entanglement on the degree of freedom of polarization, assisted by
the cross-Kerr effect in circuit QED. The principle of constructing
N-type level artificial molecule has been analyzed. Coupling to
superconducting molecule induces the effective cross-Kerr effect
between two TLRs. We make use of the effective cross-Kerr effect to
implement a QND measurement on Fock state of microwave photons in
TLRs. The QND measurement system plays a key role which can check
the polarization parity of photons in the process of entanglement
concentration. With this device, our ECP has high efficiency and it
maybe has good applications in quantum communication and quantum
computation.

\section*{ACKNOWLEDGMENT}

This work is supported by the National Natural Science Foundation of China under Grant No. 61675028, the National High Technology Research and Development Program of China under Grant No. 2013AA122902, and  the
Fundamental Research Funds for the Central Universities under Grant
No. 2015KJJCA01.


\begin{thebibliography}{1}

\bibitem{CHBennet1993} C. H. Bennett, G. Brassard, C. Crepeau, R. Jozsa, A. Peres, and W. K. Wootters, Teleporting an unknown quantum state via dual classical and Einstein-Podolsky-Rosen channels, Phys. Rev. Lett. \textbf{70}, 1895 (1993).

\bibitem{CHBennet1992} C. H. Bennett and S. J. Wiesner, Communication via one- and two-particle operators on Einstein-Podolsky-Rosen states, Phys. Rev. Lett. \textbf{69}, 2881 (1992).

\bibitem{superdense} X. S. Liu, G. L. Long, D. M. Tong, and F. Li, General scheme for superdense coding between multiparties, Phys. Rev. A \textbf{65}, 022304 (2002).

\bibitem{AKEkert} A. K. Ekert, Quantum cryptography based on Bell's theorem, Phys. Rev. Lett. \textbf{67}, 661 (1991).

\bibitem{bbm92} C. H. Bennett, G. Brassard, and N. D. Mermin, Quantum cryptography without Bell's theorem, Phys. Rev. Lett. \textbf{68}, 557 (1992).


\bibitem{lixhqkdpra} X. H. Li, F. G. Deng, and H. Y. Zhou, Efficient quantum key distribution over a collective noise channel, Phys. Rev. A \textbf{78}, 022321 (2008).


\bibitem{MHillery} M. Hillery, V. Bu\v{z}ek, and A. Berthiaume, Quantum secret sharing, Phys. Rev. A \textbf{59}, 1829 (1999).


\bibitem{longliupra} G. L. Long and X. S. Liu, Theoretically efficient high-capacity quantum-key-distribution scheme, Phys. Rev. A \textbf{65}, 032302 (2002).

\bibitem{FGDeng2003} F. G. Deng, G. L. Long, and X. S. Liu, Two-step quantum direct communication protocol using the Einstein-Podolsky-Rosen pair block, Phys. Rev. A \textbf{68}, 042317 (2003).

\bibitem{wangchQSDC} C. Wang,  F. G. Deng,  Y. S. Li,  X. S. Liu, and G. L. Long, Quantum secure direct communication with high-dimension quantum superdense coding, Phys. Rev. A \textbf{71}, 044305 (2005).







\bibitem{faithful} X. H. Li, F. G. Deng, and H. Y. Zhou, Faithful qubit transmission against collective noise without ancillary qubits, Appl. Phys. Lett. \textbf{91}, 144101 (2007).

\bibitem{DFS1}  Z. D. Walton, A. F. Abouraddy, A. V. Sergienko, B. E. A. Saleh, and M. C. Teich, Decoherence-free subspaces in quantum key distribution, Phys. Rev. Lett. \textbf{91}, 087901 (2003).

\bibitem{DFS2} J. C. Boileau, D. Gottesman, R. Laflamme, D. Poulin, and R. W. Spekkens, Robust polarization-based quantum key distribution over a collective-noise channel, Phys. Rev. Lett. \textbf{92}, 017901 (2004).

\bibitem{DFS3} J. C. Boileau, R. Laflamme, M. Laforest, and C. R. Myers, Robust quantum communication using a polarization-entangled photon pair, Phys. Rev. Lett. \textbf{93}, 220501 (2004).

\bibitem{EPP1}C. H. Bennett, G. Brassard, S. Popescu, B. Schumacher, J. A. Smolin, and W. K. Wootters, Purification of noise entanglement and faithful teleportation via noisy channels, Phys. Rev. Lett. \textbf{76}, 722--725 (1996).

\bibitem{EPP2}Y. B. Sheng and F. G. Deng, Deterministic entanglement purification and complete nonlocal Bell-state analysis with hyperentangledment, Phys. Rev. A \textbf{81}, 032307 (2010).

\bibitem{EPP3}Y. B. Sheng and F. G. Deng, One-step deterministic polarization-entanglement purification using spatial entanglement, Phys. Rev. A \textbf{82}, 044305 (2010)

\bibitem{EPP4}X. H. Li, Deterministic polarization-entanglement purification using spatial entanglement, Phys. Rev. A \textbf{82}, 044304 (2010).

\bibitem{EPP5}B. C. Ren, F. F. Du, and F. G. Deng, Two-step hyperentanglement purification with the quantum-state-joining method, Phys. Rev. A \textbf{90}, 052309 (2014).





\bibitem{CHBennett1996} C. H. Bennett, H. J. Bernstein, S. Popescu, and B. Schumacher, Concentrating partial entanglement by local operations, Phys. Rev. A \textbf{53}, 2046 (1996).

\bibitem{ZZhao014301} Z. Zhao, J. W. Pan, and M. S. Zhan, Practical scheme for entanglement concentration, Phys. Rev. A \textbf{64}, 014301 (2001).

\bibitem{TYamamoto012304} T. Yamamoto, M. Koashi, and N. Imoto, Concentration and purification scheme for two partially entangled photon pairs, Phys. Rev. A \textbf{64}, 012304 (2001).

\bibitem{YBSheng2008} Y. B. Sheng, F. G. Deng, and H. Y. Zhou, Nonlocal entanglement concentration scheme for partially entangled multipartite systems with nonlinear optics, Phys. Rev. A \textbf{77}, 062325 (2008).

\bibitem{SBose194} S. Bose, V. Vedral, P. L. Knight, Purification via entanglement swapping and conserved entanglement, Phys. Rev. A \textbf{60}, 194 (1999).

\bibitem{YBSheng012307} Y. B. Sheng, L. Zhou, S. M. Zhao, and B. Y. Zheng, Efficient single-photon-assisted entanglement concentration for partially entangled photon pairs, Phys. Rev. A \textbf{85}, 012307 (2012).

\bibitem{FGDeng022311} F. G. Deng, Optimal nonlocal multipartite entanglement concentration based on projection measurements, Phys. Rev. A \textbf{85}, 022311 (2012).

\bibitem{BCRen012302} B. C. Ren, F. F. Du, and F. G. Deng, Hyperentanglement concentration for two-photon four-qubit systems with linear optics, Phys. Rev. A \textbf{88}, 012302 (2013).

\bibitem{BCRen6547} B. C. Ren and G. L. Long, General hyperentanglement concentration for photon systems assisted by quantum-dot spins inside optical microcavities, Opt. Express \textbf{22}, 6547-6561 (2014).

\bibitem{BCRen16444} B. C. Ren and G. L. Long, Highly efficient hyperentanglement concentration with two steps assisted by quantum swap gates, Sci. Rep. \textbf{5}, 16444  (2015).

\bibitem{XHLi125201} X. H. Li and S. Ghose, Hyperconcentration for multipartite entanglement via linear optics, Laser phys. Lett. \textbf{11}, 125201 (2014).

\bibitem{XHLi3550} X. H. Li and S. Ghose, Efficient hyperconcentration of nonlocal multipartite entanglement via the cross-Kerr nonlinearity, Opt. Express \textbf{23}, 3550-3562 (2015).

\bibitem{Lixhpraecp} X. H. Li and S. Ghose, Hyperentanglement concentration for time-bin and polarization hyperentangled photons, Phys. Rev. A \textbf{91}, 062302 (2015).

\bibitem{WangECPOE15} T. J. Wang, L. L. Liu, R. Zhang, C. Cao, and C. Wang, One-step hyperentanglement purification and hyperdistillation with linear optics, Opt. Express \textbf{23}, 009284 (2015).

\bibitem{WangECPAP16} C. Cao, T. J. Wang, S. C. Mi, R. Zhang, C. Wang, Nonlocal hyperconcentration on entangled photons using photonic module system, Ann. Phys. \textbf{369}, 128--138 (2016).

\bibitem{CWangECPPRA11} C. Wang, Y. Zhang and G. S. Jin, Entanglement purification and concentration of electron-spin entangled states using quantum-dot spins in optical microcavities, Phys. Rev. A \textbf{84}, 032307 (2011).

\bibitem{CWangECPPRA12} C. Wang, Efficient entanglement concentration for partially entangled electrons using a quantum-dot and microcavity coupled system, Phys. Rev. A \textbf{86}, 012323 (2012).






\bibitem{ABlais} A. Blais, R. S. Huang, A. Wallraff, S. M. Girvin, and R. J. Schoelkopf, Cavity quantum electrodynamics for superconducting electrical circuits: An architecture for quantum computation, Phys. Rev. A \textbf{69}, 062320 (2004).

\bibitem{AWallraff} A. Wallraff, D. I. Schuster, A. Blais, L. Frunzio, R. S. Huang, J. Majer, S. Kumar, S. M. Girvin, and R. J. Schoelkopf, Strong coupling of a single photon to a superconducting qubit using circuit quantum electrodynamics, Nature(London) \textbf{431}, 162 (2004).




\bibitem{ABlais2} A. Blais, J. Gambetta, A. Wallraff, D. I. Schuster, S. M. Girvin, M. H. Devoret, and R. J. Schoelkopf, Quantum-information processing with circuit quantum electrodynamics, Phys. Rev. A \textbf{75}, 032329 (2007).


\bibitem{DiCarlo} L. DiCarlo, J. M. Chow, J. M. Gambetta, Lev S. Bishop, B. R. Johnson, D. I. Schuster, J. Majer, A. Blais, L. Frunzio, S. M. Girvin, and  R. J. Schoelkopf, Demonstration of two-qubit algorithms with a superconducting quantum processor, Nature \textbf{460}, 240 (2009).


\bibitem{LongcircuitPRA} Y. Cao, W. Y. Huo, Q. Ai, and G. L. Long, Theory of degenerate three-wave mixing using circuit QED in solid-state circuits, Phys. Rev. A \textbf{84}, 053846 (2011).



\bibitem{Wangsuperconducting} H. Wang, M. Mariantoni, R. C. Bialczak, M. Lenander, E. Lucero, M. Neeley, A. D. O'Connell, D. Sank, M. Weides, J. Wenner, T. Yamamoto, Y. Yin, J. Zhao, J. M. Martinis, and A. N. Cleland, Deterministic entanglement of photons in two superconducting microwave resonators, Phys. Rev. Lett. \textbf{106}, 060401 (2011).


\bibitem{circuitTianlPRL} Y. Hu and L. Tian, Deterministic generation of entangled photons in superconducting resonator arrays, Phys. Rev. Lett. \textbf{106}, 257002 (2011).


\bibitem{3q} A. Fedorov, L. Steffen, M. Baur, M. P. da Silva, and A. Wallraff, Implementation of a Toffoli gate with superconducting circuits, Nature \textbf{481}, 170 (2012).


\bibitem{3q1} M. D. Reed, L. DiCarlo, S. E. Nigg, L. Sun, L. Frunzio, S. M. Girvin, and R. J. Schoelkopf, Realization of three-qubit quantum error correction with superconducting circuits, Nature \textbf{482}, 382 (2012).


\bibitem{Frederick1} F. W. Strauch, All-resonant control of superconducting resonators, Phys. Rev. Lett. \textbf{109}, 210501 (2012).


\bibitem{HuaMSR1}  M. Hua,  M. J. Tao, and F. G. Deng, Fast universal quantum gates on microwave photons with all-resonance operations in circuit QED, Sci. Rep. \textbf{5}, 9274 (2015).


\bibitem{HuaMSR2}  M. Hua,  M. J. Tao,  F. G. Deng, and G. L. Long, One-step resonant controlled-phase gate on distant transmon qutrits in different 1D superconducting resonators, Sci. Rep. \textbf{5}, 14541 (2015)


\bibitem{JQYou2005} J. Q. You and F. Nori, Superconducting circuits and quantum information, Phys. Today \textbf{58}, 42 (2005).


\bibitem{LFrunzio} L. Frunzio, A. Wallraff, D. I. Schuster, J. Majer, and R. J. Schoelkopf, Fabrication and characterization of superconducting circuit QED devices for quantum computation, IEEE Trans. Appl. Supercond. \textbf{15}, 860 (2005).


\bibitem{DISchuster} D. I. Schuster, A. A. Houck, J. A. Schreier, A. Wallraff, J. M. Gambetta, A. Blais, L. Frunzio, J. Majer, B. Johnson, M. H. Devoret, S. M. Girvin, and R. J. Schoelkopf, Resolving photon number states in a superconducting circuit, Nature (London) \textbf{445} 515 (2007).

\bibitem{AAHouck} A. A. Houck, D. I. Schuster, J. Gambetta, J. A. Schreier, B. R. Johnson, J. M. Chow, L. Frunzio, J. Majer, M. H. Devoret, S. M. Girvin, and R. J. Schoelkopf, Generating single microwave photons in a circuit, Nature (London) \textbf{449}, 328 (2007).

\bibitem{JMajer} J. Majer, J. M. Chow, J. M. Gambetta, J. Koch, B. R. Johnson, J. A. Schreier, L. Frunzio, D. I. Schuster, A. A. Houck, A. Wallraff, A. Blais, M. H. Devoret, S. M. Girvin, and R. J. Schoelkopf, Coupling superconducting qubits via a cavity bus, Nature (London) \textbf{449}, 443 (2007).


\bibitem{MHofheinz} M. Hofheinz, E. M. Weig, M. Ansmann, R. C. Bialczak, E. Lucero, M. Neeley, A. D. O'Connel, H. Wang, J. M. Martinis, and A. N. Cleland, Generation of Fock states in a superconducting quantum circuit, Nature (London) \textbf{454}, 310 (2008).

\bibitem{BRJohnson} B. R. Johnson, M. D. Reed, A. A. Houck, D. I. Schuster, Lev S. Bishop, E. Ginossar, J. M. Gambetta, L. DiCarlo, L. Frunzio, S. M. Girvin and R. J. Schoelkopf, Quantum non-demolition detection of single microwave photons in a circuit, Nature Phys. \textbf{6}, 663 (2010).


\bibitem{SRebic2009} S. Rebi\'{c}, J. Twamley, and G. J. Milburn, Giant Kerr nonlinearities in circuit quantum electrodynamics, Phys. Rev. Lett. \textbf{103}, 150503 (2009).

\bibitem{SKumarPRB2010} S. Kumar and D. P. DiVincenzo, Exploiting Kerr cross nonlinearity in circuit quantum electrodynamics for nondemolition measurements, Phys. Rev. B \textbf{82}, 014512 (2010).


\bibitem{YHu} Y. Hu, G. Q. Ge, S. Chen, X. F. Yang, and Y. L. Chen, Cross-Kerr-effect induced by coupled Josephson qubits in circuit quantum electrodynamics, Phys. Rev. A \textbf{84}, 012329 (2011).


\bibitem{GKirchmair} G. Kirchmair, B. Vlastakis, Z. Leghtas, S. E. Nigg, H. Paik, E. Ginossar, M. Mirrahimi, L. Frunzio, S. M. Girvin, and R. J. Schoelkopf. Observation of quantum state collapse and revival due to the single-photon Kerr effect, Nature (London) \textbf{495} 205 (2013).


\bibitem{ICHoi} I. C. Hoi, A. F. Kockum, T. Palomaki, T. M. Stace, B. Fan, L. Tornberg, S. R. Sathyamoorthy, G. Johansson, P. Delsing, and C. M. Wilson, Giant cross-Kerr effect for propagating microwaves induced by an artificial atom, Phys. Rev. Lett. \textbf{111}, 053601 (2013).

\bibitem{ETHolland} E. T. Holland, B. Vlastakis, R. W. Heeres, M. J. Reagor, U. Vool, Z. Leghtas, L. Frunzio, G. Kirchmair, M. H. Devoret, M. Mirrahimi and R. J. Schoelkopf, Single-photon-resolved cross-Kerr interaction for autonomous stabilization of photon-number states, Phys. Rev. Lett. \textbf{115}, 180501 (2015).

\bibitem{DMPozar} D. M. Pozar, \emph{Microwave Engineering}, (Addison-Wesley, MA, 1993).

\bibitem{JMHao} J. M. Hao, Y. Yuan, L. X. Ran, T. Jiang, J. A. Kong, C. T. Chan, and L. Zhou, Manipulating electromagnetic wave polarizations by anisotropic metamaterials, Phys. Rev. Lett. \textbf{99}, 063908 (2007).


\bibitem{DRSolli1} D. R. Solli, C. F. McCormick, R. Y. Chiao, and J. M. Hickmann, Photonic crystal polarizers and polarizing beam splitters, J. Appl. Phys. \textbf{93}, 9429 (2003).


\bibitem{RPTorres} R. P. Torres and M. F. Chtedra, Analysis and design of a two-octave polarization rotator for microwave frequency, IEEE Trans. Antennas Propag. \textbf{41}, 208 (1993).


\bibitem{MKLiu} M. K. Liu, Y. B. Zhang, X. H. Wang, and C. J. Jin, Incident-angle-insensitive and polarization independent polarization rotator, Opt. Express \textbf{18}, 11990 (2010).


\bibitem{JKoch} J. Koch, T. M. Yu, J. M. Gambetta, A. A. Houck, D. I. Schuster, J. Majer, A. Blais, M. H. Devoret, S. M. Girvin, and R. J. Schoelkopf, Charge-insensitive qubit design derived from the Cooper pair box, Phys. Rev. A \textbf{76}, 042319 (2007).

\bibitem{JSiewert} J. Siewert, R. Fazio, G. M. Palma, and E. Sciacca, Aspects of qubit dynamics in the presence of leakage, J. Low Temp. Phys. \textbf{118}, 795 (2000).


\bibitem{AImamoglu1997} A. Imamo\v{g}lu, H. Schmidt, G. Woods, and M. Deutsch, Strongly interacting photons in a nonlinear cavity, Phys. Rev. Lett. \textbf{79}, 1467 (1997).


\bibitem{DFWalls} D. F. Walls and G. J. Milburn, \emph{Quantum Optics} (Springer-Verlag, Berlin, 1994).



\end{thebibliography}
\end{document}